\documentclass[11pt,a4paper]{article}
\usepackage{arxiv}
\usepackage[utf8]{inputenc} 
\usepackage[T1]{fontenc}    
\usepackage{hyperref}       
\usepackage{url}            
\usepackage{booktabs}       
\usepackage{amsfonts}       
\usepackage{nicefrac}       
\usepackage{microtype}      
\usepackage{lipsum}		
\usepackage{graphicx}
\usepackage{natbib}
\usepackage{doi}
\usepackage{epstopdf}
\usepackage{chngcntr}

\usepackage{multirow}
\usepackage{color,soul}
\usepackage{setspace}
\title{Photothermal effects in small gold nanorod aggregates for therapeutic applications}

\date{} 					

\author{{Dheeraj Pratap$^{a,}$}\thanks{Corresponding author's email: dheeraj.pratap@csio.res.in},  Ram Krishna Shah$^{b}$, Sameer Khandekar$^{b}$, Sanjeev Soni$^{a}$ \\ \\
	$^{a}$Biomedical Applications Group, CSIR-Central Scientific Instruments Organization, Sector-30C, Chandigarh-160030\\
	$^{b}$Department of Mechanical Engineering, Indian Institute of Technology Kanpur, Kanpur-208016, India\\
}

\renewcommand{\headeright}{}
\renewcommand{\undertitle}{}

\begin{document}
\maketitle

\begin{abstract}
Proximity of the metal nanoparticles enhance the plasmonic coupling and shifts the resonance. This article presents a numerical study of the photothermal effect in aggregates of small gold nanorods considering the ordered as well as random aggregates. In the ordered aggregates, there is lateral coupling which causes blueshifts in the plasmonic resonance, while in the random aggregates there are redshifts in the plasmonic resonance. The plasmon response of latter could be tailored up to the second infrared biological therapeutic window. It has been observed that the aggregates show higher absorption power and therefore, higher temperature rise compared to the single gold nanorod or monodispersive nanorods. The absorption resonance peak position of the random aggregate depends on the incident and polarization angles of the incident light. The aggregation of the nanoparticles often inherently occurs in the biological medium which affects the photothermal process. This study helps to understand the photothermal heating of nanoparticle aggregates and the use of the optimal light source concerning the absorption peak of the aggregates suspension for therapeutic uses.  
\end{abstract}


\section{Introduction}
Metal nanoparticles have applications in several areas like plasmonic photothermal treatment of tumor~\cite{nam2009ph,choi2011tumor,pattani2015role,li2008ultra,wang2013comparison}, drug delivery~\cite{adeyemi2015evaluation,janib2010imaging,babu2013nanoparticle}, imaging~\cite{lin2010direct,wiederrecht2004near,syed2017three,wang2020phase}, nanofluidics~\cite{donner2011plasmon,gomez2017ag}, nanophotonics~\cite{dionne2012plasmonics,fang2015nanoplasmonic}, sensors~\cite{tang2014novel,chen2019ultrahigh,lee2014stretchable}, catalysis~\cite{roldan2013metal,scholten2012transition,yan2010transition}, to name a few.  Verity of nanoparticles like nanosphere~\cite{soni2015experimental}, nanorod~\cite{chen2017photothermal}, nanoring~\cite{larsson2007sensing}, nanocage~\cite{skrabalak2008gold}, nanocube~\cite{fan2008epitaxial}, triangular nanoplate~\cite{apyari2018towards}, nanostar~\cite{dutta2019gold}, nanodumble~\cite{wang2019efficient} etc. are reported in literature. These nanoparticles are prepared from several plasmonic metals like gold, silver, copper and bimetallic~\cite{soni2015experimental,apyari2018towards,ruparelia2008strain,aioub2017platinum}. For all such metallic candidates, gold is highly preferable due to its chemical inertness and ease of surface functionalization with a verity of surface ligands, such as proteins and antibodies, by forming stable bonding pairs. In the case of biological applications, the nanoparticles get aggregated within the medium. In some cases, an induced aggregation also can be done inside the biological medium. Sun et al. showed the salt-induced nanoparticle aggregation could enhance the photothermal efficiency and photoacoustic imaging~\cite{sun2016salt}. Using the nuclear-targeted gold nanorods extremely low intensity near-infrared (NIR) activated photothermal therapy can be realised~\cite{pan2017nuclear}. The gold nanorods naturally accumulate in the lysosomes of the cells after reaching inside it and create clusters of the nanorods where a strong plasmonic coupling between gold nanorods exist~\cite{ungureanu2011light,shen2013targeting}. Chen et al. reported the photothermal therapy of a single cell where gold nanorods had been accumulated within the cell~\cite{chen2017photothermal}. In this case, the plasmonic resonance of the monodispersive nanorods was at 800 nm and they used a pulsed laser of 800 nm wavelength for photothermal effect. In the aggregate of the gold nanorods, there should be a resonance shift due to coupling. The use of the corresponding resonance wavelength laser could be better option but there was no mention of laser which have peak at the shifted wavelength.  Xie et al. reported a numerical investigation on absorption and plasmonic coupling in the near-field limit of the spherical nanoparticle clusters numerically~\cite{xie2019dependent}. Gu et al. used monodispersive gold nanorods for plasmonic photothermal therapy and showed that if the concentration of nanorods was increased then there was an increase of plasmonic coupling~\cite{gu2018effect}.  

Upon irradiation, the nanoparticles generate heat due to photothermal effect. Almost all applications work on the concept of plasmonic resonance. When the nanoparticle absorbs the light then it gets heated and act as a nano source of heat. Heat generation in such nanoparticles is affected by several parameters like constituent materials, particles size and shape, surrounding medium,  irradiation intensity as well as wavelength, angle of incidence etc. Plasmonic heating of two gold nanospheres have been reported by Baffou et al.~\cite{baffou2010nanoscale}. The optical properties for spherical nanoparticles have been extensively reported in literature~\cite{lerme1998optical,cottancin2006optical,kreibig2013optical}. The shape of the nanoparticles will also affect the plasmonic coupling, and hence the resonance shift in the aggregate. In literature,  heat generation characteristics of plasmonic nanoparticle aggregates is not reported. For achieving the plasmonic resonance peak at a higher wavelength in the NIR range, the gold nanorod is a good candidate because of ease of synthesis and tunability of plasmon response. One can tune the plasmonic resonance peak by varying the aspect ratio of the nanorod; however increasing the aspect ratio increases the length of the nanorod. Very large size nanoparticles are not desirable for biomedical \textit{in vivo} applications. For biological applications, there is a high possibility that the nanoparticles turn into aggregates because of interaction with the complex body fluids. Therefore, we consider the aggregates of small gold nanorods for our current study. Most of the synthesised gold nanorods have the geometry of a cylinder capped with hemispheres at both ends. When a particle is geometrically symmetric, like a sphere, then its optical properties and effect of heating could be computed analytically. However, for the case of random shape particle aggregate, analytical solutions are difficult to obtain. Further, asymmetric particle's interaction with light also depends on the polarization of the light. 

In this article, we consider small sized gold nanorods and their ordered and random aggregates in water to study the optical properties and photothermal heating numerically. The organization of the article is as follows: Section-2 describes the theoretical background of optical scattering and heating by the nanorods aggregate; Section-3 presents the numerical simulation of photothermal effect/interaction the gold nanorods and their aggregates and the results. In the final Section-4, we conclude our results.
\section{Scattering by nanoparticle aggregates}
When light interacts with a metallic nanoparticle then the free electrons start oscillating along with the electric field of the light. This oscillation of the free electrons is called localised surface plasmon (LSP)~\cite{maier2007plasmonics}. When the frequency of light matches with the natural frequency of the nanoparticles then a resonance occurs which is known as localised surface plasmon resonance (LSPR)~\cite{maier2007plasmonics}. The oscillation of electrons inside the nanoparticle due to the rapid changing electric field behaves as a varying local electron current which produces the resistive (Joule) heating in the nanoparticle~\cite{baffou2013thermo,baffou2017thermoplasmonics}. This localised heat gets diffused into the surrounding medium and the nanoparticle acts as a nanosource of heat.

Consider light wave of frequency $\omega$ and free space wavelength $\lambda_0$ having electric and magnetic field components \textbf{E$_0$}(\textbf{r},t) and \textbf{B$_0$}(\textbf{r},t) interacts with the an assembly or aggregate of nanoparticles situated in a surrounding medium (water) of permittivity $\varepsilon_s$. After interacting with assembly, some parts of light gets extinct via absorption and scattering. Geometrical factors which quantify the absorption, scattering and extinction are called  absorption cross-section ($\sigma_{\mathrm{abs}}$), scattering cross-section ($\sigma_{\mathrm{sca}}$) and extinction cross-section ($\sigma_{\mathrm{ext}}$) respectively. These quantities are defined as~\cite{mishchenko2002scattering},
\begin{equation}
\sigma_{\mathrm{abs}}=\frac{W_{\mathrm{abs}}}{P_0}, \quad \sigma_{\mathrm{sca}}=\frac{W_{\mathrm{sca}}}{P_0}, \quad \sigma_{\mathrm{ext}}=\sigma_{\mathrm{abs}}+\sigma_{\mathrm{sca}},	
\label{eq:sigma_opt_cross_sec}
\end{equation} 
%
where $W_\mathrm{abs}$ and $W_\mathrm{sca}$ are the absorbed and scattered energy by the nanoparticle aggregate and $P_0$ is the irradiance (power per unit area) of the incident light. The absorbed energy by the nanoparticle aggregate is given by the following expression,
\begin{equation}
W_\mathrm{abs}=\int_V w(\textbf{r}) d\textbf{r}
\end{equation}
where $w(\textbf{r})=(\varepsilon_s \omega/2)\mathrm{Im}(\varepsilon_\omega)|\textbf{E}(\textbf{r})|^2$ is the dissipative power density within the nanoparticle aggregate and $\varepsilon_\omega$ is the permittivity of the material of the nanoparticle. Due to the absorbed power/dissipative power density, the nanoparticle acts as a heat source and dissipates/diffuses the heat to the surrounding medium (e. g. water or tissue). The main mode of heat transfer in plasmonics is diffusion. To get the temperature field, $T$, of the nanoparticle and its surrounding medium we have to solve the heat diffusion equation, 
\begin{equation}
\rho_\omega c_{p,\omega} \partial_t T - \kappa_\omega \nabla^2 T = w,
\label{eq:diff_inside}
\end{equation}
inside the nanoparticle, where $\rho_\omega$, $c_{p,\omega}$, $\kappa_\omega$ and $T$ are density, specific heat capacity at constant pressure and thermal conductivity of the material of the nanoparticle, and local temperature respectively; and 
\begin{equation}
\rho_s c_{p,s} \partial_t T - \kappa_s \nabla^2 T = 0,
\label{eq:diff_outside}
\end{equation}
outside the nanoparticle, where $\rho_s$, $c_{p,s}$ and $\kappa_s$ are density, specific heat capacity at constant pressure and thermal conductivity of the surrounding medium (water)~\cite{baffou2013thermo,baffou2017thermoplasmonics}.

For the symmetric geometries, like a sphere, the analytical solutions for the optical cross-sections, dissipative power and temperature field can be obtained under the imposed boundary condition. However, analytical solutions are difficult for irregular shapes or geometries. For random shaped nanoparticles, numerical modelling tools like boundary element method (BEM)~\cite{de2002retarded}, Green's dyadic method (GDM)~\cite{martin1995generalized,girard2008shaping}, discrete dipole approximation (DDA)~\cite{girard2005near,yurkin2007discrete}, multiple elastic scattering of multipole expansions (MESME) ~\cite{myroshnychenko2008modelling,de1999interaction}, finite difference time domain (FDTD)~\cite{radhakrishnan2014calculation,mohsin2018probing}, finite element method (FEM)~\cite{bachelier2008fano,alizadeh2019investigating} etc. are used for solution. In the present work, we have used COMSOL Multiphysics$^{\textregistered}$~\cite{multiphysics1998}, a FEM based commercial tool to model the scattering, absorption and heating of nanorod aggregates. 
\begin{figure}[h]
\centering 
\includegraphics[width=0.9\textwidth]{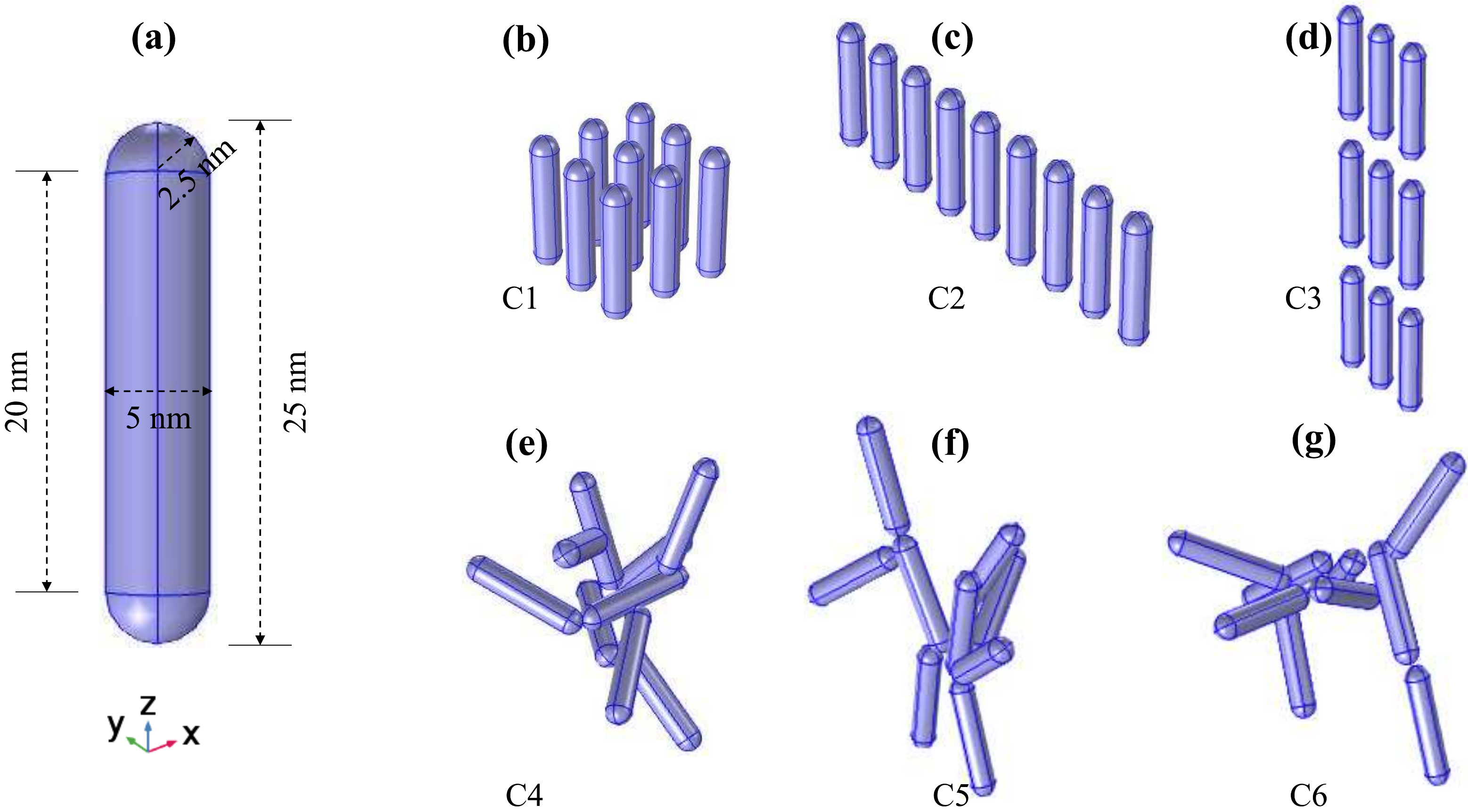}
\caption{Schematic of (a) a gold nanorod of length 25 nm and diameter 5 nm. Default three-dimensional views of geometrical structures of (b-d) ordered aggregates made of nine nanorods C1, C2 and C3; and (e-g) random aggregates made of nine nanorods C4, C5 and C6. The projected views of the aggregates in the $xy$, $yz$ and $zx$ planes are shown in the Appendix Fig.~\ref{fig:cl_projections}.}
\label{fig:cl_schematic}
\end{figure} 

Figure~\ref{fig:cl_schematic} delineates the geometries of a gold nanorod and its aggregates. The central part of the nanorod has a cylindrical shape of length 20 nm and diameter 5 nm. At both ends of the 20 nm cylinder, there are hemispheres of diameter 5 nm. The total length of the gold nanorod is 25 nm and the aspect ratio is 5. The considered nanorod dimension and aspect ratio is not restricted, one could take nanorods of other size and aspect ratio too. Choosing the larger single particle may increase the scattering and reduce the heating. Here, our main objective is to study the plasmonic coupling and heating in aggregates of small nanorods, as nanoparticles of size below 50 nm are preferred for therapeutic applications. There are only nine nanorods assumed in each aggregate. In the first three aggregates the nanorods are arranged in an ordered manner (C1, C2 and C3), and in the other three aggregates the nanorods are oriented randomly (C4, C5 and C6). The projected views of all six considered aggregates in the $xy$, $yz$ and $zx$ planes are provided in the Appendix Fig.~ \ref{fig:cl_projections}. We choose only nine nanorods to construct the ordered and random  aggregates design so that aggregates should not become very large and nanorods could be arranged symmetrically also for comparison with the random assembly configurations. We did not consider a linear chain of nine nanorods arranged in end-to-end (longitudinal coupling) because in such a long chain large redshift is obvious, and practically such a singular arrangement will occur rarely in reality. All nanorods (nine) are identical in each assembly so that their optical properties and photothermal effects could be compared. Choosing a larger size of single particle may increase the scattering and reduce the heating. For instance, a nanosphere of 100 nm diameter gives SPR in the mid-visible range and at this scale; the scattering phenomena dominates over the absorption. If we want to tune the LSPR of the aggregates of nanospheres then there would be too much scattering because of the large size. For the biomedical \textit{in vivo} application, smaller nanoparticles are preferred. The small nanorod of aspect ratio 5 itself gives the LSPR at near-infrared (NIR) wavelength range which helps to tune the LSPR of the aggregates for higher wavelength in the second biological window easily. We have chosen three ordered assemblies (C1, C2 and C3) for comparison. For practical cases, such ordered nature of assembly would not normally occur. Naturally, there could be countless type of assemblies possible, and even in a single assembly, the nanorods could have an infinite number of orientations. Such a huge possibility of the random system is difficult to model in a single study, therefore, we have considered here only three aggregates of the random orientation of nine nanorods to understand the photothermal behaviour of the same. 

To model the nanorod and their aggregates to study the photothermal effects in the COMSOL$^{\textregistered}$, we first carried out the scattering of a gold nanosphere of 100 nm diameter in water, heated by the light of intensity 1 mW/$\mu$m$^2$ at a wavelength 530 nm as a benchmark study, which was numerically reported by  Baffou et al. using the BEM method~\cite{baffou2010nanoscale}, and later reproduced by Chen et. al. using COMSOL Multiphysics$^{\textregistered}$~\cite{chen2012nanosecond}. Results of the benchmark study are given in the Appendix Fig.~\ref{fig:ns_heating}. Our results are very closely matching with the results of Chen~\cite{chen2012nanosecond} and have slight deviation compared with the results of Baffou~\cite{baffou2010nanoscale}. A slight variation of results is due to the fact that Baffou used the BEM and Chen used the FEM based COMSOL$^{\textregistered}$ and both approaches have different working principle. 
\begin{figure}[h]
\centering
\includegraphics[width=1\textwidth]{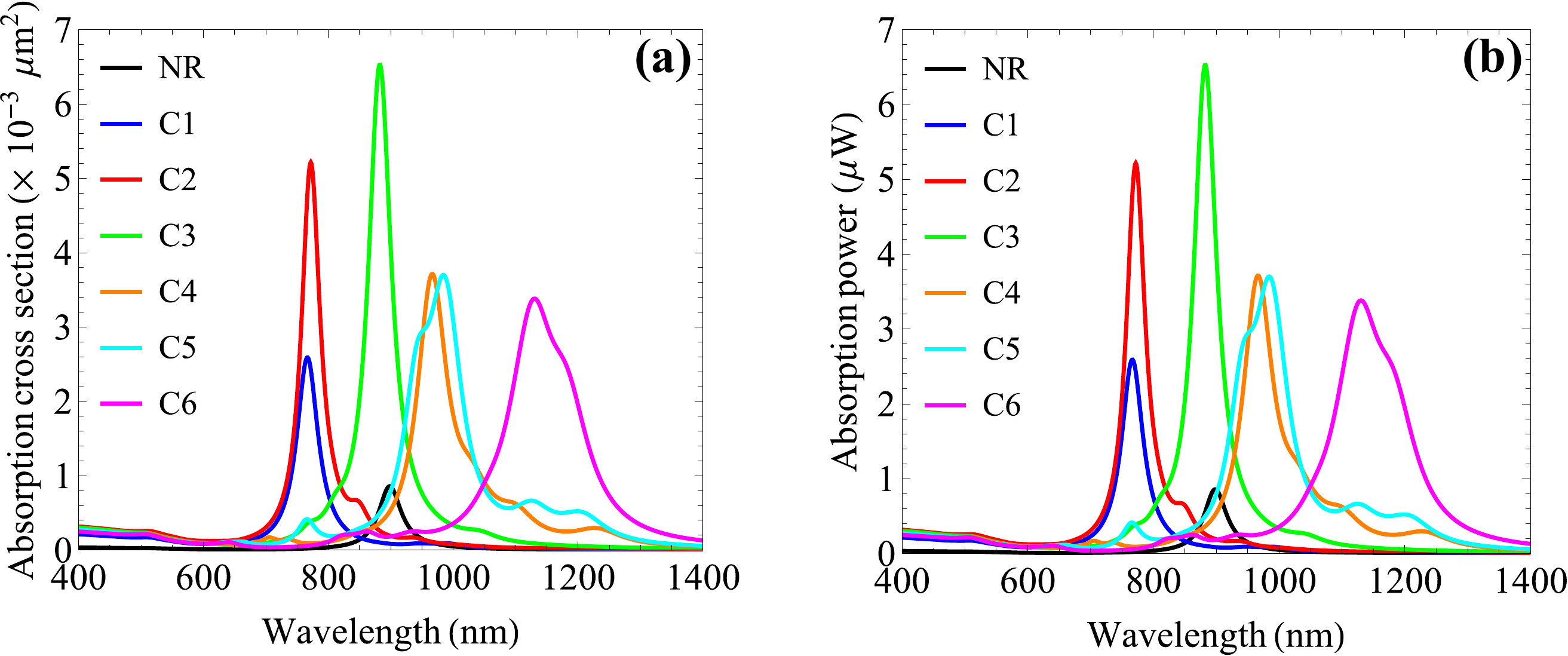}
\caption{(a) Absorption cross section and (b) absorption power of the gold nanorod aggregates C1, C2, C3, C4, C5 and C6 dispersed in water. For comparison, the absorption cross section and absorption power of a single gold nanorod is also plotted along with the aggregates. The nanorod has length 25 nm and  diameter 5 nm.}
\label{fig:cl_abs_cross_sec_pow}
\end{figure} 

The optical scattering of the nanorod and their aggregates were modelled in COMSOL$^{\textregistered}$ by considering that the nanorods assembly is situated in a spherical domain of water. At the outer side of the water domain, a perfectly matched layer (PML) was used to avoid reflection from the water boundary.  The incident unpolarised electromagnetic field was assumed in background with propagation vector $\textbf{k}=k_0(\mathrm{sin}(\theta)\mathrm{cos}(\phi),\mathrm{sin}(\theta)\mathrm{sin}(\phi),\mathrm{cos}(\theta))$ at $\theta=45^{\circ}$ and $\phi=45^{\circ}$. The standard dispersive permittivity of the gold was taken from Johnson and Christy~\cite{johnson1972optical}, and the permittivity of the surrounding medium, i.e., water, was kept constant = 1.777. We computed the optical cross-sections of the gold nanorod and their ordered and random aggregates. Figure~\ref{fig:cl_abs_cross_sec_pow} shows the computed optical cross-sections and absorption power of the gold nanorod and their aggregates. In all ordered aggregates the inter-particle distance is fixed to be 5 nm and in the random aggregates the inter-particle distance is variable and the nanorods are not penetrating each other. The strongest LSPR peaks are listed in Table~\ref{tab:SPR_abs_pow_dT}. The longitudinal LSPR peak of a single gold nanorod is at 898 nm. When the nanoparticles are ordered in an aggregate (C1, C2 and C3) then there is a blueshift in the LSPR due to the lateral coupling within the nanorods. There is the longitudinal coupling in aggregate C3 but still, there is a blueshift in LSPR because the lateral coupling dominates over longitudinal coupling. As we had considered only a fixed inter-particle distance in the ordered aggregates because it was well known that decreasing (or increasing) inter-particle distance would increase the coupling, therefore, there would be more blueshift (or redshift) in the LSPR peaks. In the case of the random orientation of the nanorods in aggregates (C4, C5 and C6) there are redshifts in the LSPR peaks. Practically, there would be random distribution in the aggregates, therefore, it is good that we can tune the LSPR in the second biological therapeutic window, as shown in Fig.~\ref{fig:cl_abs_cross_sec_pow}. The total absorbed power by an aggregate is also higher than a single nanorod. 
\begin{table}[h]
\centering
\caption{The strongest LSPR wavelength, total absorption power and computed maximum temperature change of the gold nanorod aggregates.}
\begin{tabular}{| c | c | c | c |}
\hline 
Sample & Strongest LSPR peak (nm) & Max. abs. power ($\mu$W) & Max. $\Delta$T (K)\\
\hline \hline
NR & 898 & 0.86 & 18.4 	\\ \hline  
C1 & 767 & 2.59 & 23.3 	\\ \hline  
C2 & 772 & 5.21 & 35.4 	\\ \hline  
C3 & 883 & 6.53 & 44.4 	\\ \hline  
C4 & 967 & 3.71 & 27.3 	\\ \hline  
C5 & 984 & 3.70 & 26.2 	\\ \hline  
C6 & 1130 & 3.38 & 22.5 \\ \hline  
\end{tabular}
\label{tab:SPR_abs_pow_dT}
\end{table}

\begin{figure}[h]
\centering
\includegraphics[width=0.7\textwidth]{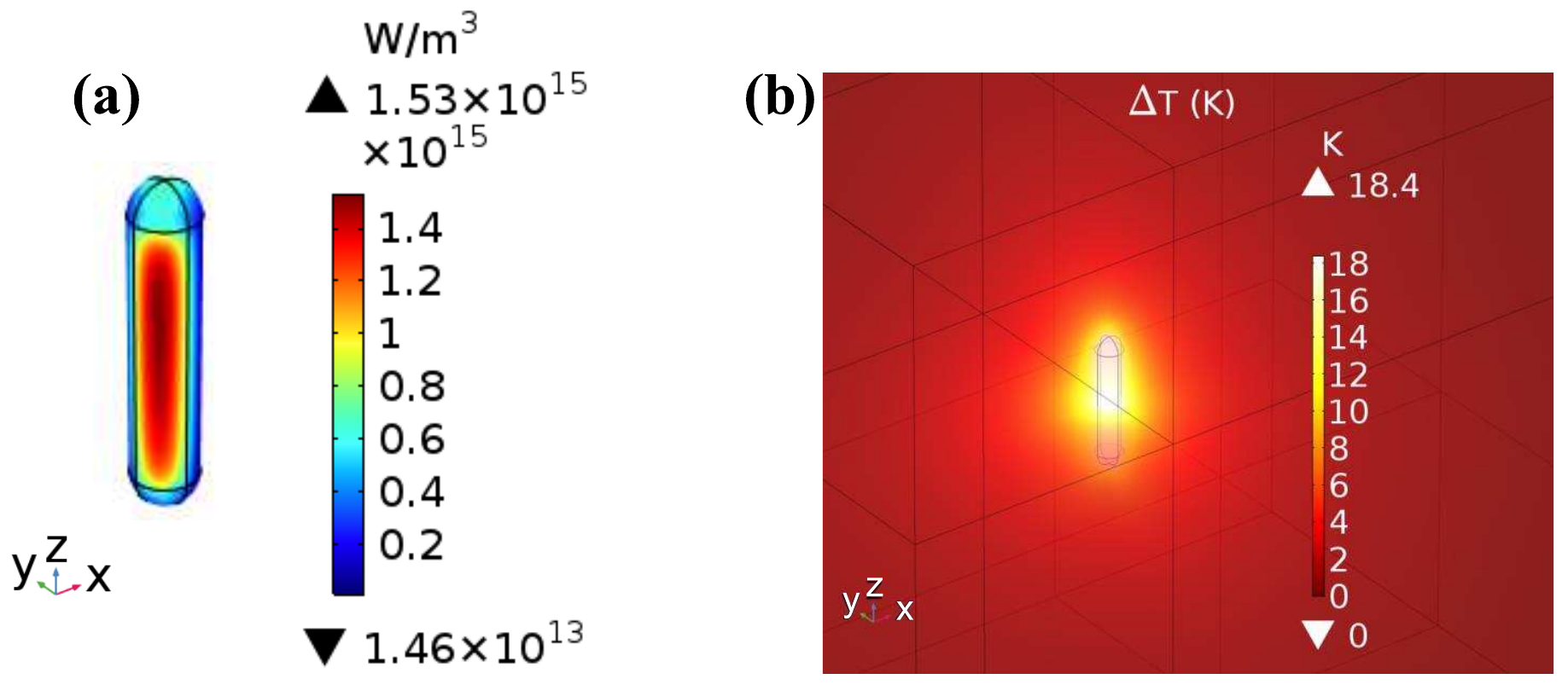}
\caption{(a) Surface plot of the total absorption power density and (b) increases temperature field at the $x = 0$, $y = 0$ and $z = 0$ planes of a single gold nanorod in water domain.}
\label{fig:nr_pow_den_dT}
\end{figure}

To study the resistive heating in the gold nanorod and aggregates we had considered a cubical domain of water where nanorod or aggregate was situated. At the top face of the cubical water domain, facing along the z-direction, we made the optical port for input light. We considered a general wave vector at $\theta=45^{\circ}$ and $\phi=45^{\circ}$ propagating towards negative z-direction and the unpolarised electric field was perpendicular to the incident wave propagation vector. A PML layer at the outer surface of the water domain was taken. A constant temperature boundary condition ambient temperature, ($T_{\mathrm{amb}}=300.15$K) was used at the outermost surface of the computational domain. We used the nanorod or aggregate as a heat source where the already calculated absorbed power was used as total dissipated heat energy inside the nanorod or aggregate. We solve the stationary state condition to get the maximum raised temperature of the system. Density, specific heat capacity and thermal conductivity of gold were $19.3 \times 10^3$ kg/m$^3$, 129 J/(kg.K) and 318 W/(m.K), respectively. For the water density, specific heat capacity and thermal conductivity were 1000 kg/m$^3$, $4.2 \times 10^3$ J/(kg.K) and 0.6 W/(m.K) respectively. Figure~\ref{fig:nr_pow_den_dT} shows the surface plot of total power dissipation density and the increased temperature field ($\Delta T$) at the $x = 0$, $y = 0$ and $z = 0$ planes of a single gold nanorod in the water domain. We can see that the maximum temperature increase was 18.4 K for a single gold nanorod. 

\begin{figure}[h]
\centering
\includegraphics[width=1\textwidth]{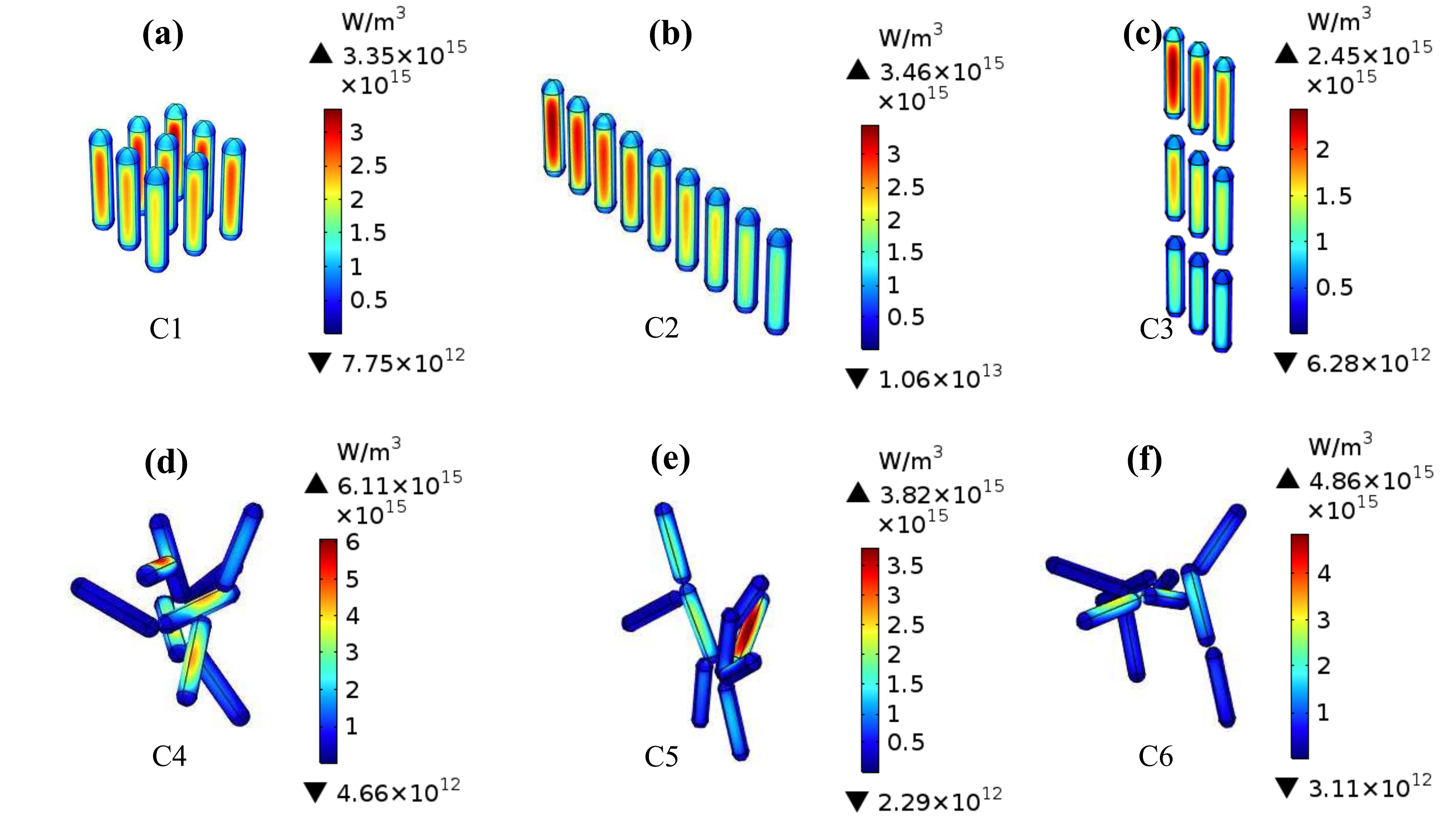}
\caption{Surface plots of the absorption power density of (a-c) the ordered aggregates C1, C2 and C3; and (b) random aggregates C4, C5 and C6 of gold nanorods.}
\label{fig:cl_pow_den}
\end{figure}

Figure~\ref{fig:cl_pow_den} shows the surface plot of the total absorption power density of the ordered and random aggregates of the gold nanorods. Here, due to the three-dimensional distribution of the nanorods, the slice plot (at $x = 0$, $y = 0$ and $z = 0$ planes) was not visible properly, therefore, we used a surface plot of the absorbed power density on the nanorods' assembly. In the surface plot, there were finite minimum and maximum value of the absorption power density. Since the direction of the propagation vector was at $\theta=45^\circ$ and $\phi=45^\circ$, therefore, in the ordered aggregates a relative gradient (shadow) of the absorption power could be seen from one nanorod to another. If propagation would have at $\theta=0^\circ$ and $\phi=0^\circ$, then there would have no gradient (shadow) of the absorption power density from particle to particle. The case of random aggregate is comparatively complex than the ordered aggregates. So we can't see proper shadowing effects on the random aggregates. In the case of aggregates, the absorption power density is higher than the single gold nanorod (Fig.~\ref{fig:nr_pow_den_dT} (a)). 

\begin{figure}[h!]
\centering
\includegraphics[width=1\textwidth]{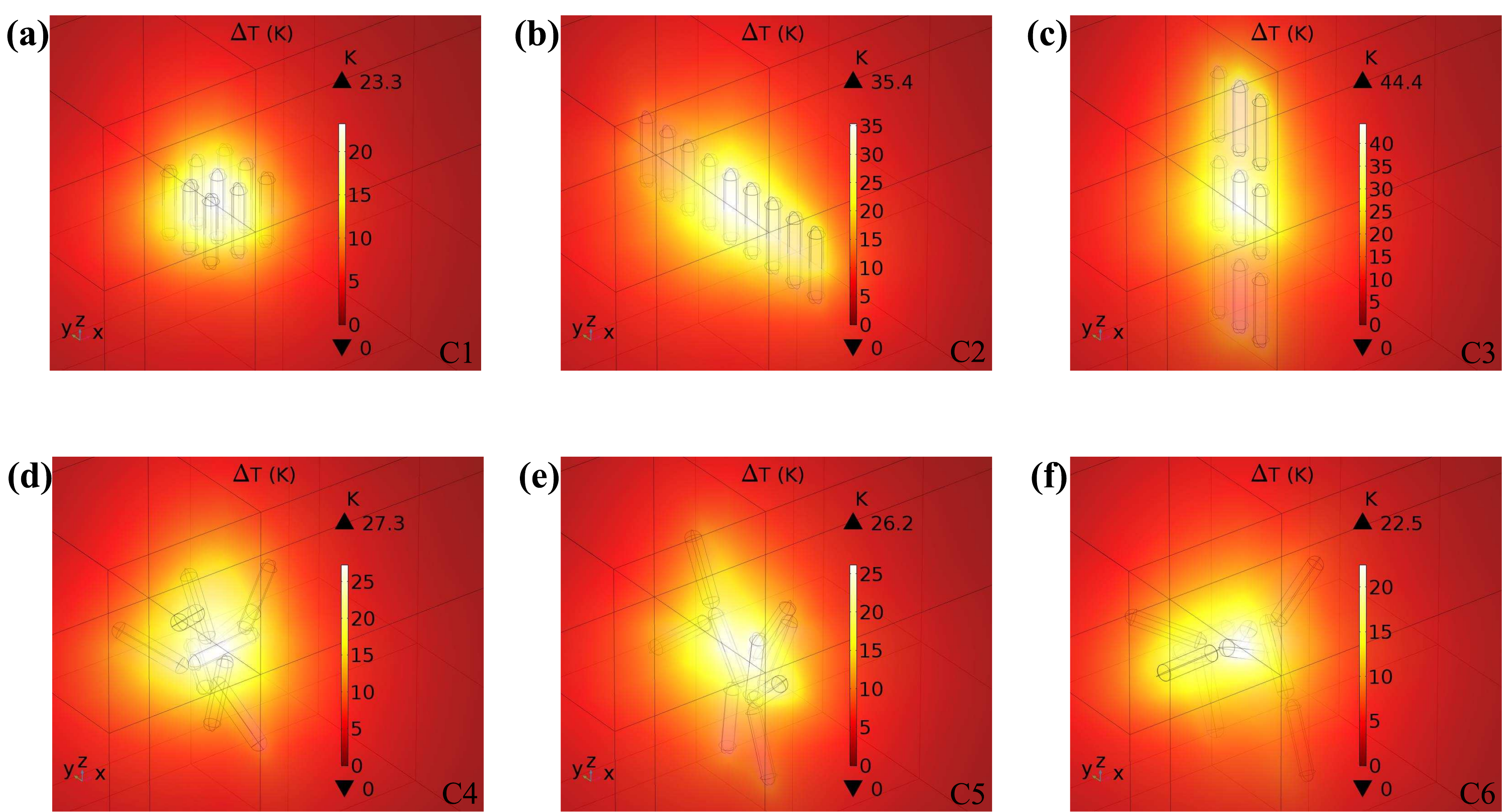}
\caption{Temperature field profile ($\Delta T$) at the $x = 0$, $y = 0$ and $z = 0$ planes of the (a-c) ordered (C1, C2 and C3) and (d-f) random (C4, C5 and C6) gold nanorod aggregates.}
\label{fig:cl_dT}
\end{figure}

The temperature rise in the ordered and random aggregates are shown in Fig.~\ref{fig:cl_dT}. We could see that aggregates got higher temperature compared to the single nanorod (monodispersive). In all six aggregates, ordered aggregates have higher temperature compared to the random aggregates, since each entity in the ordered assembly participates in the plasmonic coupling, leading to more temperature rise than the random case. However, in the case of random assembly, every particle has different coupling, some higher and some lower. One should not worry about the higher temperature rise in ordered assembly compared to the random assembly because practically ordered assembly would not be ideally possible in a solution dispersive form. At the nanoscale, aggregation of the nanorods is random, which have a higher temperature rise compared to a single nanorod (monodispersive). Since the LSPR peak for the random aggregates occurs at a larger wavelength compared to a single (monodispersive) nanorod, therefore, incident light of higher wavelength would allow deeper penetration and thus higher temperature rise in the tissue. Because of LSPR at higher wavelength, random aggregates, therapeutically have more advantages over monodispersive nanoparticles. The size of aggregation is not fixed by a few numbers of nanorods or nanoparticles, it could be anything. However, through proper control over the experimental parameters, the aggregate size and hence its LSPR peak could be controlled. If we could control the size of the aggregates then we could decide the light source of desired higher wavelength for therapeutic use. The current photothermal case of small gold nanorod aggregate is more oriented toward the \textit{in vivo} therapeutic use in tumor domain thermal therapy etc., but it is not limited to other uses also. For example, the case of plasmonic photothermal application in antimicrobial uses and acne treatments are not much limited over the size of aggregation and these applications have more advantages to increase the higher surface temperature and destroying the bacteria.  

\begin{figure}[h!]
\centering
\includegraphics[width=1\textwidth]{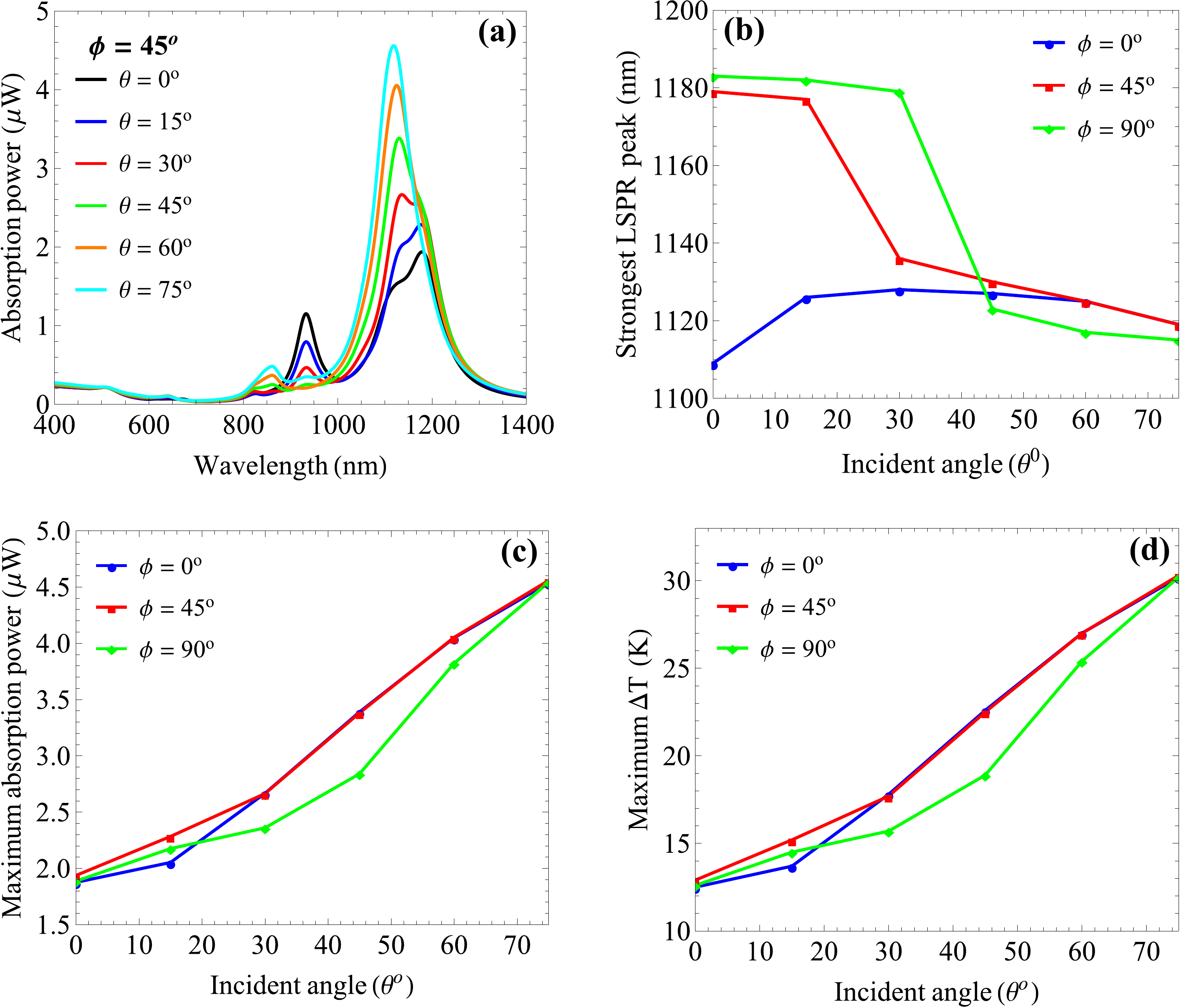}
\caption{Variation of absorption characteristic and temperature rise of gold nanorod aggregate C6 for various incident and polarization angles of the light. (a) Absorption power of the aggregate with incident wavelength at different incident angle $\theta$ for polarization angle $\phi=45^{\circ}$, (b) the strongest LSPR peaks, (c) maximum absorption power at the strongest LSPR peak and (d) maximum temperature change $\Delta T$ (K) at illumination with strongest LSPR wavelength with respect to the indecent angle at different polarization angle.}
\label{fig:cl_angle_dependent}
\end{figure}
For most of the practical purpose, the incident light would be randomly polarized and aggregates would be randomly oriented. The exact experimental condition could not be obtained in the numerical studies. However, we carried out the angle-dependent study of the light absorption and heat generation by the random aggregate system C6 only. The angle $\theta$ is the inclination or incident angle from the $z$-axis of the light wavevector and $\phi$ is the polarization angle from the $x$-axis of the electric field. We fixed the polarization angle $0^{\circ}$, $45^{\circ}$ and $90^{\circ}$ and varied the incident angle from $0^{\circ}$ to $75^{\circ}$ in the interval of $15^{\circ}$ and computed the light absorption and heat generation. Figure~\ref{fig:cl_angle_dependent} shows the results of the incident light's angle-dependent study for the random aggregate C6. Figure~\ref{fig:cl_angle_dependent}(a) presents the total absorption power of the aggregate with respect to the incident wavelength at a fixed polarization angle $45^{\circ}$ and at different incident angles. We can see that as the  wavevector inclination angle $\theta$ increases, there is a slight monotonic blueshift in the strongest LSPR peak. This happens due to the reason that the nanorods in the aggregates are randomly oriented and at a different angle there would be variation in the plasmonic coupling within the particles. For the monodispersive nanorods, the blueshift could not happen by varying the incident and polarization angles but only there would be a change in the absorption amplitude. Variation in the absorption peaks for aggregates with incident and polarization angle implies that in the real experimental system there would be a broadening in the LSPR peaks. Figure~\ref{fig:cl_angle_dependent}(b) shows the strongest LSPR peaks with respect to the incident angle at fixed polarization $0^{\circ}$, $45^{\circ}$ and $90^{\circ}$. Here we see how the strongest LSPR peak position varies, indicating  the broadening of the peaks in experimental conditions. Figure~\ref{fig:cl_angle_dependent}(c) and Figure~\ref{fig:cl_angle_dependent}(d) show the maximum absorption power and corresponding maximum temperature change at the strongest LSPR peaks with respect to the incident angle at different polarization angle $0^{\circ}$, $45^{\circ}$ and $90^{\circ}$ respectively. We could notice that the polarization did not change absorption power and temperature rise much due to the random nature of the aggregate. One should note that all these computed observations are just for one particular aggregate configuration, however, these results would naturally vary for other configurations of the aggregates. By fixing the experimental conditions one could get the well defined LSPR peak for aggregates suspension in the solution. 
\section{Conclusion}
We conclude that aggregation of the small gold nanorods causes a blueshift of LSPR wavelength in ordered aggregates and a redshift of the LSPR wavelength in the random aggregates. Even if there was either a blueshift or redshift in the LSPR peak of any aggregate, yet, there was a higher temperature rise in that aggregate as compared to the single nanorod (monodispersive). The angle-dependent study in the random aggregates shows that there is a slight monotonic blueshift in the LSPR peak which would cause broadening of the LSPR peak of aggregates suspension in an actual experiment.  The size and structure of the aggregation would depend on the experimental or local environmental or chemical conditions with optimization of the experimental conditions, the desired type of a stable aggregate could be obtained. We could say that the aggregation of the gold nanorods or any shape of metallic nanoparticles would have clear advantages also over monodispersive form, at least wherever there would be a limitation of the size of the nanoparticle and need of higher wavelength plasmonic coupling and greater temperature rise. For example, in the case of plasmonic photothermal therapy of tumor tissue to get higher light penetration depth,  antimicrobial applications, and plasmonic photothermal treatment of acne, such aggregates will produce more heat than their monodispersive counterparts. 
\section*{Acknowledgements}
D.P. acknowledges the CSIR-India for the Nehru Science Postdoctoral Research Fellowship number HRDG/CSIR-Nehru PDF/EN, ES \& PS/EMR-I/04/2019. D.P. and S.S. acknowledge the support of the CSIR-CSIO Chandigarh for hosting the research.

\appendix
\newpage
\counterwithin{figure}{section}  
\section{Aggregate Projections}
\begin{figure}[b!]
 \centering 
\includegraphics[width=0.8\textwidth]{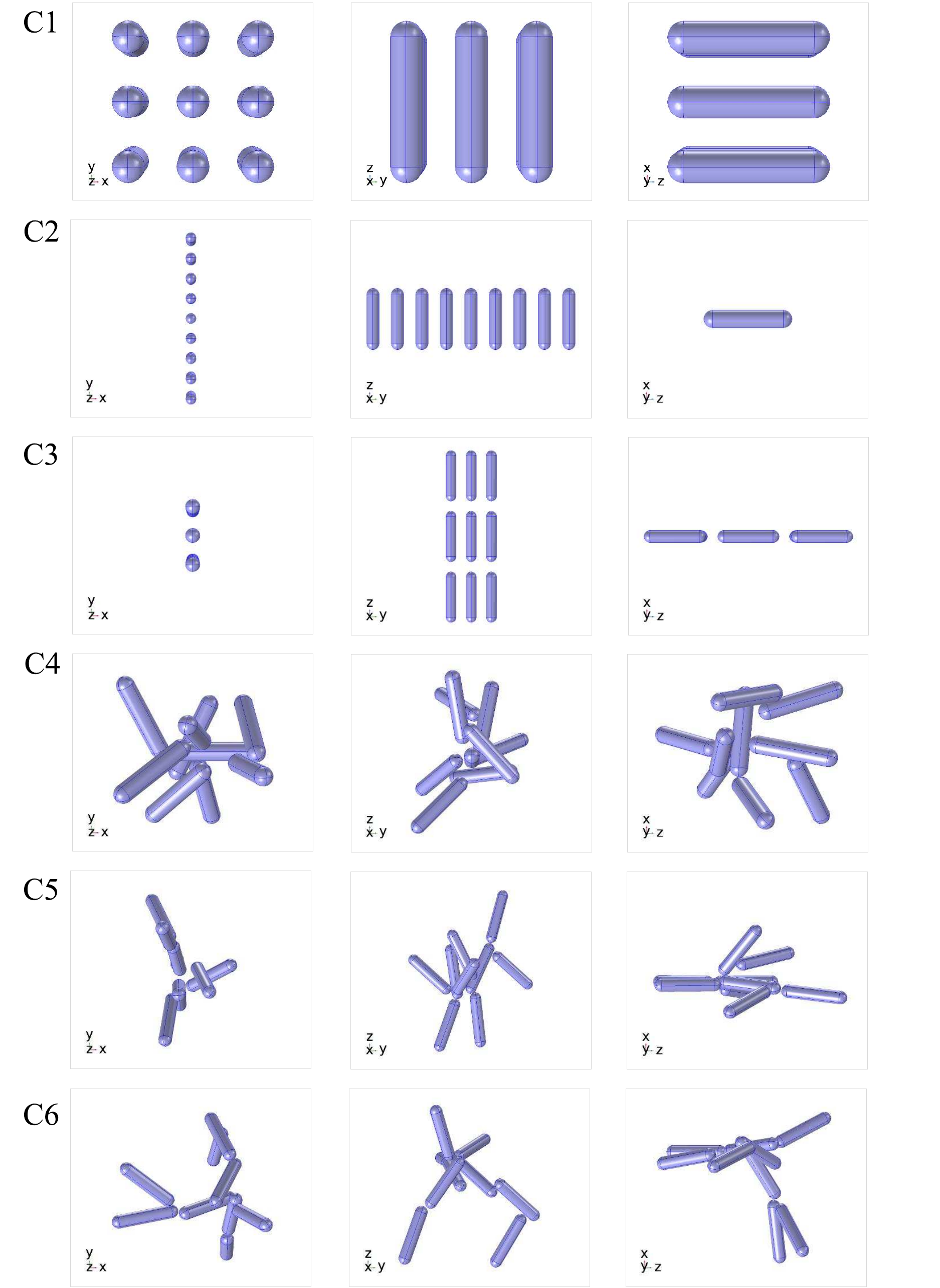}
\caption{Projected views of aggregates C1, C2, C3, C4, C5 and C6 in the $xy$, $yz$ and $zx$ planes used for computation. Each nanorod has length 25 nm and diameter 5 nm.}
\label{fig:cl_projections}
\end{figure}

All considered six aggregates, ordered as well as random, are three-dimensional entities. The geometric structure projections in the $xy$, $yz$ and $zx$ planes are shown in the Fig.~\ref{fig:cl_projections}. 

\newpage
\section{Model validation}
\begin{figure}[h]
 \centering 
\includegraphics[width=1\textwidth]{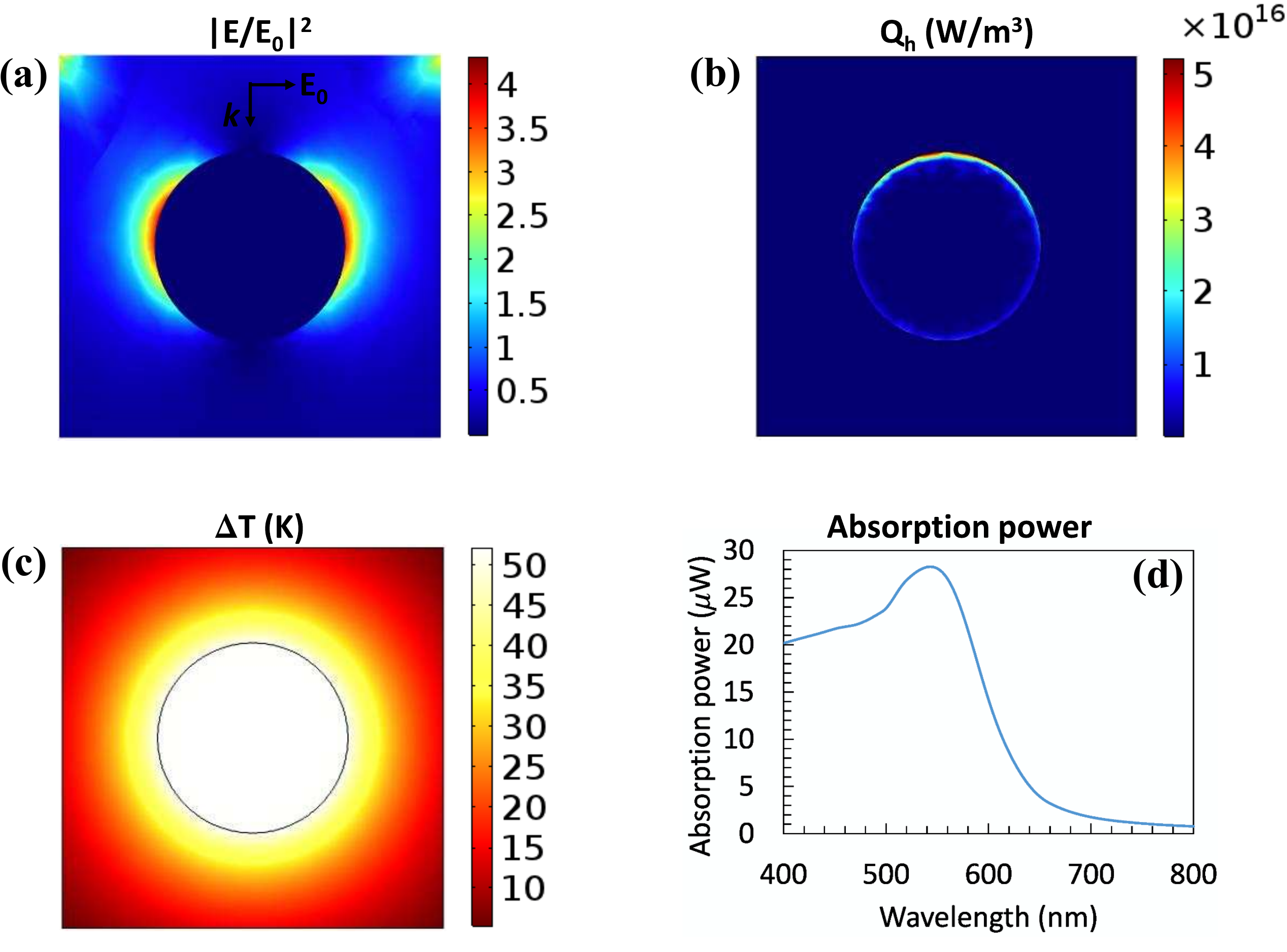}
\caption{Comparison validation of algorithm (a) normalised electric field intensity, (b) absorbed power density, (c) temperature field and  (d) total absorbed power of a gold nanosphere of 100 nm diameter at incident wavelength 530 nm and irradiance 1 mW/$\mu$m$^2$.}
\label{fig:ns_heating}
\end{figure}

Heat generation study of a 100 nm gold nanosphere situated in water was carried out as a benchmark problem. The permittivity of the gold was from Johnson and Christy~\cite{johnson1972optical}. All values of other parameters had been kept identical as used by G. Baffou et al.~\cite{baffou2010nanoscale}. Figure~\ref{fig:ns_heating} shows the computation of the normalised electric field intensity, energy density, temperature evolved in the nanosphere at incident wavelength 530 nm.. The total absorption power of the nanosphere from 400 nm to 800 nm is also shown. The irradiance of the light was 1 mW/$\mu$m$^2$. In our case, the maximum raised temperature was 52.2 $^\circ$C which was 55 $^\circ$C in Baffou case and 52.5 $^\circ$C in X. Chen et al.~\cite{chen2012nanosecond}. The maximum absorbed power was obtained 28.27 $\mu$W at 543 nm while the maximum absorbed power reported by Baffou was 20.5 $\mu$W at 530 nm and by Chen 21.45 $\mu$W at 540 nm.  

\bibliographystyle{unsrt}
\bibliography{references}
\end{document}